\documentclass{article}
\usepackage[english]{babel}
\usepackage[utf8]{inputenc}
\usepackage{csquotes}
\usepackage{johd}
\usepackage{url}
\usepackage{amsmath, amssymb}
\usepackage{hyperref}
\usepackage{graphicx}
\usepackage{subfigure}
\usepackage{multirow}
\usepackage{booktabs}
\usepackage{geometry}
\usepackage{indentfirst}
\title{Evaluation of Multi-indicator And Multi-organ Medical Image Segmentation Models}

\author{Qi Ye, Lihua Guo}

\date{} %leave blank

\begin{document}

\maketitle

\begin{abstract} 
\noindent In recent years, ``U-shaped'' neural networks featuring encoder and decoder structures have gained popularity in the field of medical image segmentation. Various variants of this model have been developed. Nevertheless, the evaluation of these models has received less attention compared to model development. In response, we propose a comprehensive method for evaluating medical image segmentation models for multi-indicator and multi-organ (named MIMO). MIMO allows models to generate independent thresholds which are then combined with multi-indicator evaluation and confidence estimation to screen and measure each organ. As a result, MIMO offers detailed information on the segmentation of each organ in each sample, thereby aiding developers in analyzing and improving the model. Additionally, MIMO can produce concise usability and comprehensiveness scores for different models. Models with higher scores are deemed to be excellent models, which is convenient for clinical evaluation. Our research tests eight different medical image segmentation models on two abdominal multi-organ datasets and evaluates them from four perspectives: correctness, confidence estimation, Usable Region and MIMO. Furthermore, robustness experiments are tested. Experimental results demonstrate that MIMO offers novel insights into multi-indicator and multi-organ medical image evaluation and provides a specific and concise measure for the usability and comprehensiveness of the model. \\
Code: \url{https://github.com/SCUT-ML-GUO/MIMO}

\end{abstract}


\section{Introduction}

  \noindent Due to the development of deep learning methodologies, computer vision technology has become a crucial component in medical image segmentation tasks. The ``U-shaped'' encoder-decoder architecture, particularly the U-Net \cite{3,7} model, has yielded significant success in the domain of medical semantic segmentation. However, owing to the increasing novelty and complexity of medical image segmentation tasks, the vanilla U-Net may not suffice to meet the requirements. Consequently, researchers have sought to address these challenges by improving the vanilla U-Net in various directions and with varying complexities. As a result, numerous models, such as Attention u-net \cite{6}, nnU-Net \cite{5}, and Transunet \cite{8}, have been developed, and have achieved state-of-the-art outcomes in diverse medical image segmentation tasks and challenges \cite{12,14,15}. These advancements have greatly facilitated the evolution of medical image segmentation.

  Despite the rapid evolution of medical image segmentation models, the evaluation methods for these models in the context of specific tasks (e.g., multiple indicators and multiple organs) remain limited. Traditional evaluation methods (e.g., Dice coefficient, Hausdorff distance) primarily assess the accuracy of the model's prediction outcomes. While high accuracy is desirable, it is insufficient to establish the practical clinical utility of the model. Since every model is prone to erroneous predictions, clinicians still encounter issues of reliability when deploying these models in clinical practice.

  Investigations into model reliability have gained momentum, with the expected calibration error \cite{25} and maximum calibration error \cite{25} being the widely used approaches. Lower values of these metrics suggest that the model's confidence scores are better aligned with the actual correctness scores. Despite predictive accuracy and confidence estimation reliability being crucial for the clinical and practical implementation of medical image segmentation models, neither of these aspects can determine the model's overall availability in isolation. Thus, Zhang et al. \cite{23} proposed a new measurement method, namely Correctness-Confidence Rank Correlation (CCRC), along with the Usable Region Estimate. This method gauges the usability of the prediction outcomes in the confidence space and performs well in single-index and single-organ segmentation tasks.

  However, in practical applications, researchers commonly employ several metrics to evaluate prediction accuracy, and medical image segmentation tasks involving multiple organs, it will be challenging when setting manual requirements for the Usable Region Estimate. Additionally, there is still no established methodology for uniformly and reasonably assessing different models in specific application scenarios, such as those involving multiple organs and indicators.

  From the perspective of comprehensiveness and unity, we propose the multi-indicator and multi-organ evaluation (MIMO) approach to evaluate medical image segmentation models. Specifically, MIMO quantifies the accuracy of multi-indicator prediction scores and the reliability of confidence estimates for the multi-organ assessment using a reasonable and unified metric. A model is deemed more usable and reliable for clinical applications only if it demonstrates better accuracy scores (e.g., larger dice coefficient and smaller Hausdorff distance) and higher confidence estimation to get a good final score (between 0 and 1). The collection of sample organs generated during the operation of the algorithm also helps developers to analyze and improve the model.

  More clinical-oriented model evaluation can promote more clinical-oriented model development, our main contributions to this work are as follows::
  
\begin{itemize}
\item We propose a novel evaluation method for multi-index and multi-organ medical image segmentation models. Allowing models to generate thresholds independently and using these thresholds to screen the sample organs after the joint ranking of predictive correctness index and confidence estimation. Then, MIMO feedbacks on the situation that each organ in each sample meets (or does not meet) the standard accurately, and it facilitates the subsequent analysis for improving the model. Furthermore, a usability and comprehensiveness score is output in an ``area'' manner to evaluate different models concisely and intuitively.
\item The independent generation of thresholds for each organ is evaluated through Bootstrapping \cite{24} under each evaluation index in the model. The accuracy level of these thresholds can be easily changed by adjusting the percentile.
\item To validate the effectiveness of the proposed method for multi-index multi-organ segmentation, we reproduced, tested, and evaluated eight different medical image segmentation models on two public datasets, BTCV \cite{12} and AMOS 2022 \cite{15}. The robustness of the proposed method is tested on the above data sets and shows a lower frequency of errors.
\end{itemize}

\section{Related Work}

\noindent \textbf{U-Net and its variants}: U-Net \cite{3,7} is a codec network that receives much concern in recent years. It consists of a symmetrical encoder-decoder structure with skip connections. The encoder utilizes convolutional and downsampling layers to extract depth features of the receptive field, while the decoder receives semantic information from the bottom of the ``U'' and restores spatial information through skip connections. Inspired by U-Net, Milletarì et al. \cite{4} designed a 3D segmentation model V-Net based on volume data and proposed the Dice loss function. Oktay et al. \cite{6} introduced an attention gate to U-Net and developed Attention u-net, which processes the feature map of the skip connection with attention gating. In contrast, another study \cite{5} focused on the universality of the U-Net model and made minimal modifications by changing the activation function and regularization strategy while systematizing the segmentation task to adapt to multiple datasets. Additionally, with Vision Transformers \cite{16} gaining attention in the field of computer vision, Transunet \cite{8} combines U-Net and Transformer to become the first medical image segmentation model using Transformer, while Swin-unet \cite{9} was born in combination with Swin Transformer block \cite{26}. Based on the fact that they can only deal with 2D data and other defects, Unetr \cite{10} and Swin unetr \cite{11} came into being, and have a great capability of learning multi-scale contextual representations and modeling long-range dependencies. The success of U-Net and its variants has demonstrated their potential for advancing medical image segmentation models.

\noindent \textbf{Evaluation method of medical image segmentation model}: The evaluation of medical image segmentation model prediction has traditionally focused on correctness indicators (e.g., Dice coefficient and HD distance), which have served as the basis for model improvement for a significant period of time. However, Guo et al. \cite{13} have highlighted a reduction in reliability despite the improved accuracy of today's model. Consequently, research on model calibration and the reliability of confidence estimates has received more attention. Dropout-based and Bayesian-based methods \cite{17,18} are common choices for confidence estimation, and the expected calibration error (ECE) \cite{25,13,21} is widely used to evaluate the calibration of prediction confidence. ECE calculates the absolute error between the correctness score and the reliability score (e.g., confidence) of each prediction sample, and outputs the mean. Similarly, the maximum calibration error (MCE) \cite{25,13} focuses on the maximum absolute error, while the Brier score \cite{22} measures the mean square error between each prediction probability and its corresponding true value label. New research \cite{19,20} suggests that the evaluation of medical artificial intelligence models should include both the correctness of prediction and the reliability of confidence. However, neither the correctness score nor the reliability score of confidence estimation alone can provide a unified perspective of the actual usability of the model. Zhang et al. \cite{23} proposed the measure of correctness-confidence rank correlation and further designed the usable region estimation (URE), which unifies the reliability of prediction correctness and confidence estimation.

   Our research aims to evaluate models of multi-indicator and multi-organ medical image segmentation scenarios. Unlike previous methods, our approach generates thresholds autonomously within the model. The set of sample organs that meet the thresholds is output and an intuitive usability and comprehensiveness score is then calculated, which brings new insights and concise metrics into the evaluation of multi-indicator and multi-organ medical image segmentation models.

\section{Methodology}
\subsection{Evaluation Metrics}

\noindent \textbf{Correctness of prediction.} The conventional approach to performance evaluation in medical image segmentation primarily centers on segmentation accuracy, which is typically evaluated using either region-based or boundary-based metrics. Region-based evaluation methods (e.g., Dice coefficient and IOU) are utilized to compare the similarity between the segmentation outcomes and the ground truth. Boundary-based evaluation methods, on the other hand, assess the differences between the segmentation result and the ground truth boundary, with the Hausdorff distance being the most common metric. In this study, we mainly employ the Dice coefficient and Hausdorff distance.

 For a given semantic class, let {$G_i$} and {$V_i$} denote the ground truth and prediction values for voxel \textit{i} and \textit{G\'} and \textit{V\'} denote ground truth and prediction surface point sets respectively. The Dice score and HD metrics are defined as
 \begin{equation}
    Dice(G,V) = \frac{2 \times \sum_{i=1}^I G_i \times V_i}{\sum_{i=1}^I G_i + \sum_{i=1}^I V_i}
    \label{eq:factorial1}
\end{equation}
 \begin{equation}
    HD(G',V') = \max \left\{ \max \limits_{g' \in G'} \min \limits_{v' \in V'} ||g'-v'||, \max \limits_{v' \in V'} \min \limits_{g' \in G'} ||v'-g'|| \right\}
    \label{eq:factorial2}
\end{equation}

\noindent \textbf{Reliability of Confidence Evaluation.} The ideal scenario in model output confidence estimation is that it should be equivalent to the true probability, such that $\hat{P}(\hat{Y} = Y)=p,\forall p \in [0,1]$.Where $\hat{Y}$ is the predicted output and \textit{Y} is the ground truth, $\hat{P}$ is the confidence level of the label output by the model. The confidence estimation calculation method is defined as $conf(n)=\frac{1}{n}\sum_{i \in n} p_i$. Where \textit{n} denotes the total sample size, $p_i$ is the confidence of the $i^{th}$ sample prediction. One concept of error correction is the difference between the expectation of confidence and accuracy. The ECE and MCE are two common methods used for prediction confidence calibration. In a per-sample level comparison, ECE is defined as $ECE = \sum_{i=1}^n \frac{1}{n}|s_i - conf_i|$.However, in high-risk applications, where reliable confidence measures are critical, the focus is usually on the worst-case deviation between confidence and accuracy, and MCE is defined as $MCE=\max \limits_{i \in \left\{1 \dots n \right\}}|s_i - conf_i|$. To ensure a fair and direct comparison between the evaluated models, we use a common practice: take the maximum value of the predicted logits (after Softmax) for each pixel in the sample, and then obtain the average value of all pixels in each channel of each test sample as the confidence of the model's own output.

\begin{figure}[H]
\centering
\includegraphics[width=0.9\textwidth]{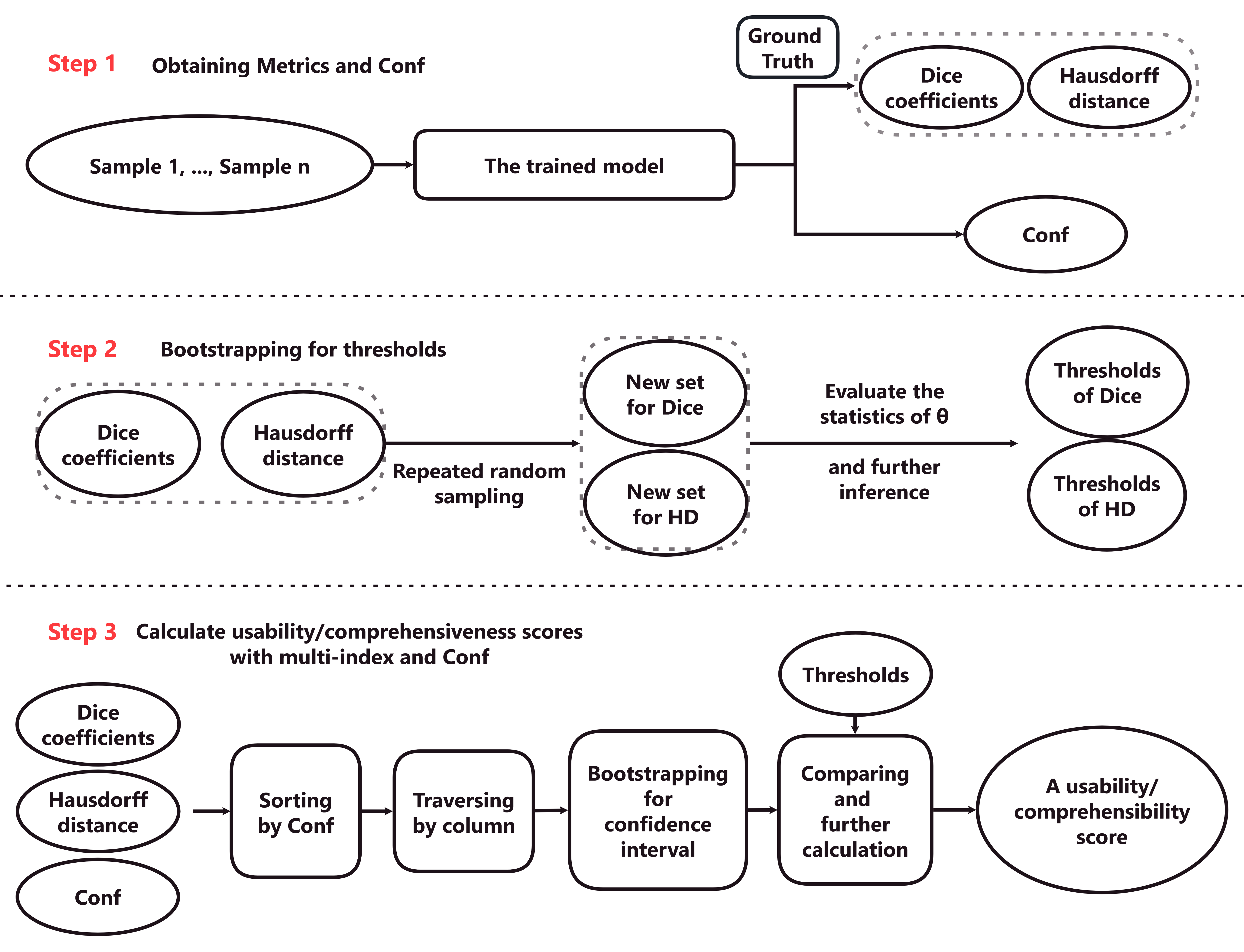}
\caption{\label{Fig.1} A diagram illustrating the overall framework of the proposed method}
\end{figure}

\subsection{The framework of MIMO}
The provided diagram in Fig.\ref{Fig.1} depicts the overall architecture of the proposed method MIMO, which can be summarized into three steps. Assuming there are \textit{n} samples and \textit{m} organs that require segmentation. 

The first step of the proposed method involves obtaining the correctness index and confidence value. The trained model is utilized to segment all test set samples and obtain the corresponding output. The Softmax activation function is applied to the output, followed by computing the mean pixel values of all test samples in each channel to derive the confidence value (Conf) for each organ. Additionally, the Dice coefficient and Hausdorff distance of each organ can be obtained as the correctness index by combining the output with the ground truth using Formula \ref{eq:factorial1} and Formula \ref{eq:factorial2}.

The second step involves utilizing the Bootstrapping algorithm \cite{24} to acquire the thresholds of different organs under each index. The Dice coefficient and Hausdorff distance are employed as two sets for repeated random sampling, and repeated \textit{B} times, with each resampled sample referred to as a Bootstrap sample. The Bootstrap sample is evaluated for the statistics of $\theta$ to obtain \textit{B} estimates, sorted in descending order, and the corresponding percentile is taken as the threshold output. For general purposes, the ``mean'' statistic is set as the default value of $\theta$ to estimate the prediction correctness.

The third step entails combining multi-index and Conf to calculate final score in multi-organ segmentation. The Dice coefficient, Hausdorff distance, confidence value, and all thresholds obtained from the previous step are inputted. Sorting the data according to Conf, traversing by column, using Bootstrapping to acquire the confidence interval and comparing the value with the threshold to determine the organ set satisfying the threshold in all samples. Finally, a usability and comprehensiveness score (between 0 and 1) can be computed by further computation. The more details are shown in section 3.3.

\subsection{Calculation of usability and comprehensiveness}
\begin{figure}[H]
\centering
\includegraphics[width=0.7\textwidth]{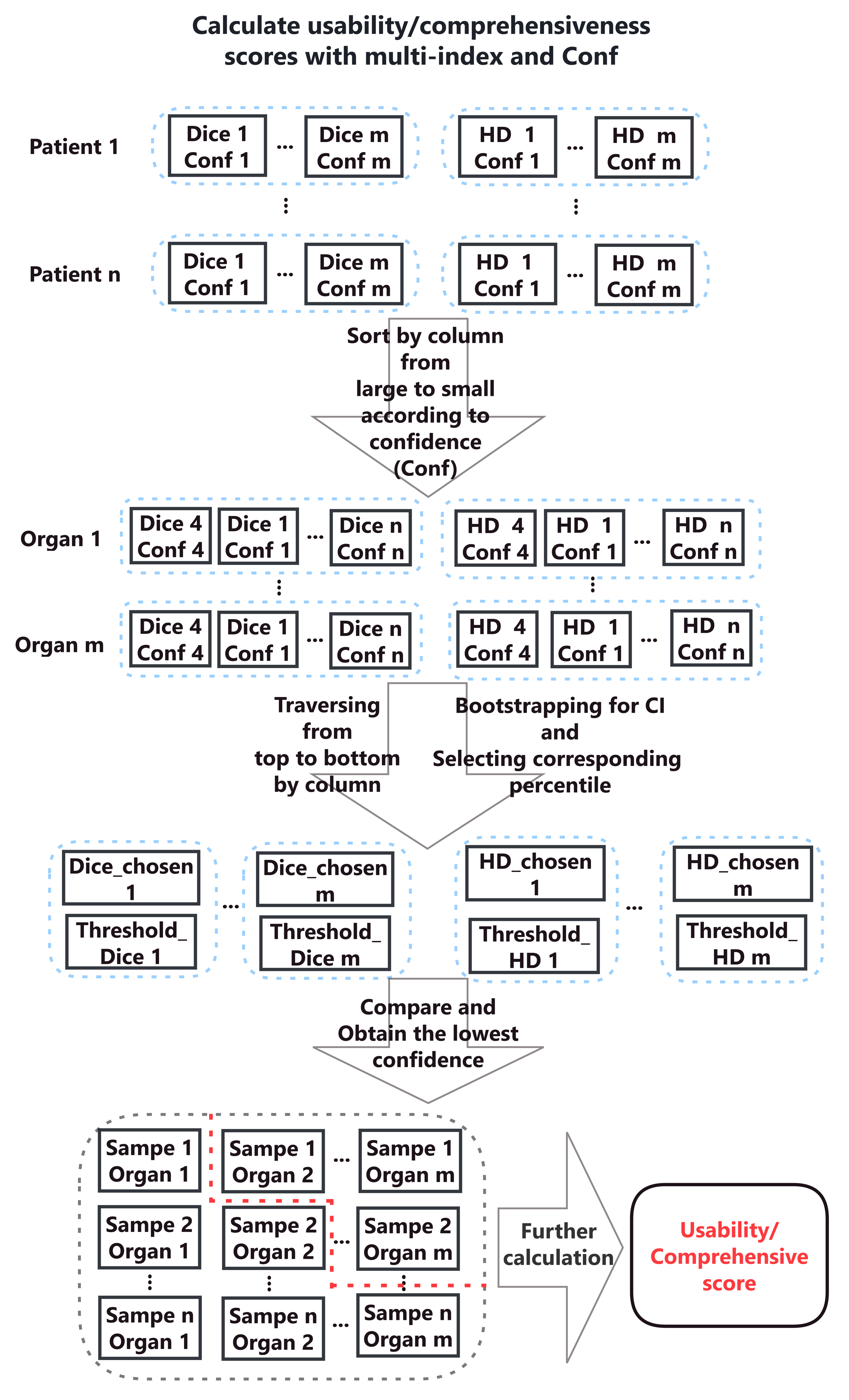}
\caption{\label{Fig.2} The concrete realization of calculating usability and comprehensiveness score. The left part of the red line in the lower left corner is the sampled organ that satisfies both Dice $\geq$ Threshold of Dice, HD $\leq$ Threshold of HD and the Conf is greater than the lowest confidence.}
\end{figure}

Fig.\ref{Fig.2} illustrates the specific implementation of calculating usability and comprehensiveness scores with multi-index and Conf in multi-organ segmentation, where each sample represents a patient case, and each column represents an organ to be segmented. Firstly, a one-to-one correlation is established between the Dice coefficient (or Hausdorff distance) results of each sample and the confidence value (Conf), and each column is sorted in descending order according to the confidence value to obtain a new set. Secondly, the new set is traversed by column. When traversing to the $i^{th}$ (\textit{i} $\leq$ column length) data of each column, the first \textit{i} Dice coefficients (or Hausdorff distance) in the previously sorted set from the initial step are taken out and Bootstrapping \cite{24} is used to calculate the confidence interval (CI). The clinically acceptable segmentation accuracy threshold should include a 95\% CI confidence interval, so a 95\% percentile is selected for the Dice coefficient, and 5\% is selected for the Hausdorff distance. Thirdly, the value of the corresponding percentile obtained by each organ under multiple indicators is compared with the threshold. The minimum confidence value satisfying the threshold is recorded, and a set of organs satisfying the threshold is generated. The traversal continues for each column until the end, and the minimum confidence value and the above set are updated. Finally, only organs satisfying both the Dice coefficient and the Hausdorff distance, and with a confidence value exceeding the minimum value, are considered qualified organs. The usability and comprehensiveness score is obtained by dividing the total number of organs that meet the conditions by the cumulative value of all the organs to be segmented in the total sample, and a higher usability and comprehensiveness score indicates a superior model. This method unifies the multiple correctness indicators, multiple organ segmentation, and reliability of confidence estimation. 

\subsection{Controllable evaluation index threshold}
During the threshold generation process, the user can manipulate the percentile (default is 50\%) to create thresholds with varying degrees of accuracy. For example, for the Dice coefficient, the larger the percentile, the stricter the corresponding demand, while the Hausdorff distance is the opposite. These thresholds will result in different sizes of target organ sets and corresponding usability and comprehensiveness scores. A model is deemed superior only if it consistently yields higher scores across multiple accuracy levels. This approach helps to avoid premature conclusions about the superiority of a given model and promotes more rigorous model development. Furthermore, the proposed method is capable of identifying instances where a model performs well on one portion of an organ in a sample but not on another. This capability enables the identification of such phenomena when comparing models, thereby facilitating further model refinement.

\section{Experiments}
\subsection{Implementation Details}
This experiment is mainly based on PyTorch\footnote{\href{https://pytorch.org/}{https://pytorch.org/}} and MONAI\footnote{\href{https://monai.io/}{https://monai.io/}}. The platform environment relied on during the experiment is Windows 10 64-bit operating system. The training and testing of models are carried out on an NVIDIA GeForce RTX 2080Ti with 11GB memory.

 The Transunet and Swin-unet models are restricted to 2D input, necessitating a data preprocessing approach that involves the conversion of 3D CT data to ``Numpy'' format, image clipping within a range of [-125, 275], normalization of each 3D image to the range [0,1], and the extraction of 2D slices from 3D volumes for training cases. The 3D volume is kept in h5 format for testing cases. During the training phase, a batch size of 24 and the popular stochastic gradient descent (SGD) optimizer with a momentum of 0.9 and weight decay of 1e-4 are utilized. Furthermore, pre-trained weights from ImageNet \cite{1} are used to initialize the model parameters.

  The remaining models are reproduced in MONAI, utilizing a batch size of 1 for model training and the AdamW optimizer \cite{2} with an initial learning rate of 0.0001 for 5000 iterations. The sliding window method is used for inference, with a 0.5 overlap between adjacent patches and a resolution of [96, 96, 96]. For the Unetr and Swin unetr models, no pre-trained weights for the Transformers backbone are used. The data preprocessing strategy involves independent normalization of each volume by rescaling the strength from the range [-1000, 1000] Hounsfield Units (HU) to [0, 1]. Additionally, all images are preprocessed to achieve an isotropic voxel spacing of 1.0 mm. The employed data enhancement strategies include random rotation of 90, 180, and 270 degrees, random flipping of axial, sagittal, and coronal positions, random scaling, and offset intensity.

\subsection{Datasets}
To assess the efficiency of our approach, we conduct experiments on eight distinct models: U-Net\cite{7}, V-Net\cite{4}, nnU-Net\cite{5}, Attention u-net\cite{6}, Transunet\cite{8}, Swin-unet\cite{9}, Unetr\cite{10} and Swin unetr\cite{11}. These models are tested on two datasets, namely BTCV\cite{12} and AMOS 2022\cite{15}, and they are evaluated from four perspectives: correctness, confidence estimation, Usable Region and MIMO.

\noindent\textbf{BTCV(CT):} The BTCV dataset comprises 50 subjects with abdominal CT scans from the MICCAI 2015 multi-atlas abdominal marker challenge. Our experiments utilize 30 abdominal CT scans with ground truth, with 20 randomly selected for training, 4 for validation, and 6 for testing. The interpreters supervised by a clinical radiologist at the Vanderbilt University Medical Center annotated 13 organs in the dataset. Each CT scan was enhanced in the portal venous phase and contained between 80 to 225 images, each with dimensions of 512 $\times$ 512 pixels and slice thickness ranging from 2.5 to 5.0 mm.

\noindent\textbf{AMOS 2022(CT):} The AMOS 2022 dataset is derived from the MICCAI 2022 multimodal abdominal multi-organ segmentation challenge, and it comprises 500 CT and 100 MRI scans collected from multiple sources, centers, and patients. Each scan features voxel-level annotations of 15 abdominal organs. Our experiments on this dataset are limited to Task 1 (CT only), utilizing 360 samples with ground truth. The training set consists of 150 cases, the validation set 30 cases, and the test set 60 cases, randomly partitioned.

\subsection{Quantitative Analysis}
\begin{figure}[htbp]
  \centering
  \subfigure{\includegraphics[width=0.6\textwidth]{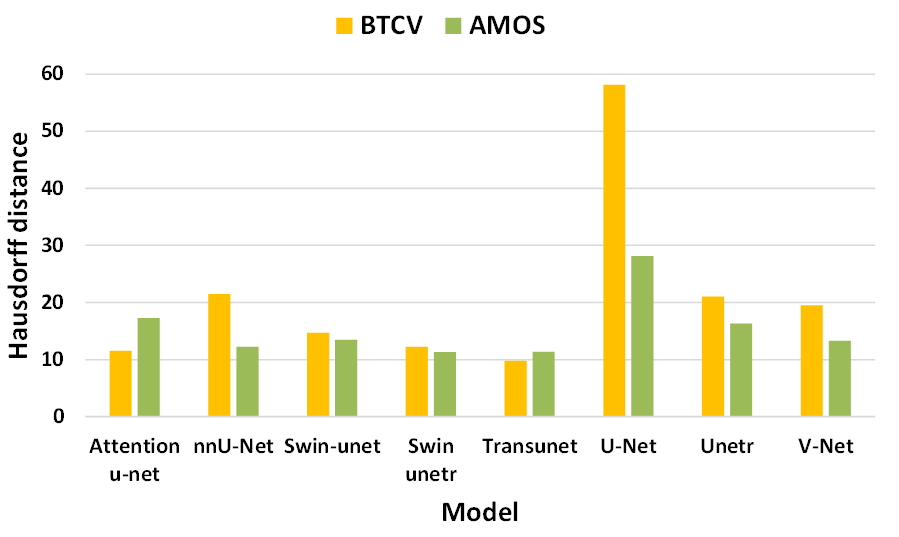}}
  \caption{The average Hausdorff distance of each model in two datasets.}
  \label{Fig.3}
\end{figure}

\noindent \textbf{Correctness.} As shown in Table\ref{tab1}, the performance of various models in two different datasets was evaluated based on the average Dice coefficient (↑). Two models, i.e., Attention u-net and Swin unetr demonstrated Dice coefficients of 0.832 and 0.838, respectively, in the BTCV dataset. In the AMOS 2022 dataset, nnU-Net and Swin unetr achieved Dice coefficients of 0.87 and 0.876, respectively. While Swin unetr performed slightly better than other models in both datasets, its margin of superiority over the second-best model was small, and it is noteworthy that the worst-performing model, in terms of Dice coefficient, was Swin-unet and U-Net in the respective datasets. Regarding average Hausdorff distance (↓), as depicted in Fig.\ref{Fig.3}, Transunet had the lowest score in the BTCV dataset and showed a significant deviation from other models. In the AMOS 2022 dataset, Swin unetr scored slightly lower than Transunet with a difference of 0.03, and ranked first, and both datasets indicate that U-Net exhibited the poorest boundary similarity. However, the ranking results of the Hausdorff distance and the Dice coefficient may conflict with each other. Such a conflict arising from the multi-index evaluation can cause difference in the model ranking, and this will make it more complicated for doctors and experts to deploy the model in clinical practice.

\begin{table}[H]
\renewcommand\arraystretch{1.5}
\centering % Label your table accordingly
\caption{\label{tab1}Average Dice coefficient and usability and comprehensiveness of each model in two datasets}
\begin{tabular}{ccccc}
\toprule[0.35mm]
\multirow{2}{*}{\textbf{Models}} & \multicolumn{2}{c}{\textbf{Average Dice coefficient}}\ & \multicolumn{2}{c}{\textbf{Usability and comprehensiveness}}\cr 
\cmidrule[0.25mm](lr){2-3} \cmidrule[0.25mm](lr){4-5}
 & BTCV & AMOS 2022 & BTCV & AMOS 2022 \\
\midrule[0.15mm]
U-Net & 0.776 & 0.798 & 0.128 & 0.238 \\
V-Net & 0.822 & 0.834 & 0.615 & 0.664 \\
nnU-Net  & 0.829 & 0.87 & 0.513 & 0.83 \\
Attention u-net & 0.832 & 0.841 & 0.705 & 0.679 \\
Transunet & 0.752 & 0.81 & 0.346 & 0.757 \\
Swin-unet & 0.714 & 0.772 & 0.295 & 0.34 \\
Unetr & 0.782 & 0.826 & 0.564 & 0.372 \\
Swin unetr & \bf{0.838} & \bf{0.876} & \bf{0.769} & \bf{0.851}\\
\bottomrule[0.35mm]
\end{tabular}
\end{table}

\begin{figure}[htbp]
  \centering
  \subfigure{\includegraphics[width=0.45\textwidth]{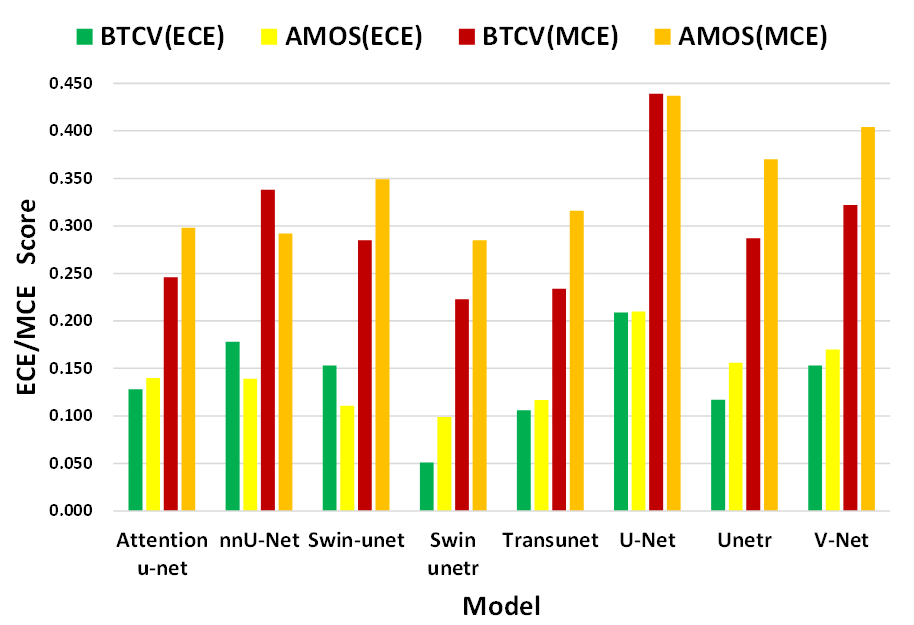}}
  \subfigure{\includegraphics[width=0.45\textwidth]{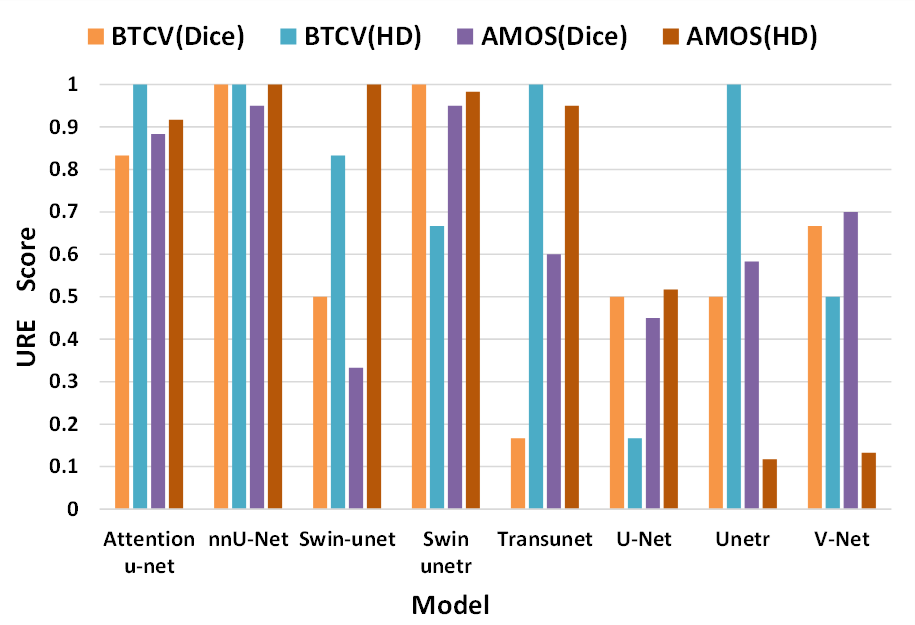}}
  \caption{Test results for ECE, MCE and Usable Region Estimate. \textbf{Left}:The test results of the expected calibration error and the maximum calibration error of each model. \textbf{Right}:The results of the Usable Region Estimate test on the gallbladder under the default percentile.}
  \label{Fig.4}
\end{figure}

\noindent \textbf{Confidence Estimation.} In regards to ECE (↓) and MCE(↓), Fig.\ref{Fig.4}(Left) reflects that Swin unetr demonstrates superior calibration levels in both datasets. Conversely, U-Net shows the highest score in both indicators and performs the worst. However, it should be noted that the ranking of other models will be different due to the consideration of ECE or MCE, and will also be affected by different data sets. Additionally, there is a phenomenon, for example, V-Net excels over Unetr in terms of correctness score but performs worse in reliability. This presents a challenge when selecting a model for clinical deployment as a decision must be made between prioritizing correctness versus reliability. Consequently, it becomes difficult to compare models straightforwardly for clinical application.

\noindent \textbf{Usable Region Estimate.} Fig.\ref{Fig.4}(Right) demonstrates the utilization of the Usable Region (↑) to evaluate the Dice coefficient and Hausdorff distance in two datasets under default percentile conditions. The gallbladder serves as an example in this case. The experimental results indicate that although the method combines both correctness and reliability, it does not address the issue of ambiguity. Furthermore, many model scores overlap, e.g., nnU-Net and Swin unetr both achieve the highest score of 1, at the same time, Swin-unet, U-Net and Unetr reach below 0.5 simultaneously. This phenomenon occurs because the sample-driven method is not suitable when a limited number of test samples exists. Ultimately, it is essential to note that diverse organs will yield different sorting results, which presents a puzzle for physicians evaluating multi-organ segmentation results in a clinical setting.

\noindent \textbf{MIMO.} The overall usability and comprehensiveness (↑) of models is shown in Table\ref{tab1}. U-Net, V-Net, nn U-Net, Attention u-net, Transunet, Swin-unet, Unetr, and Swin unetr achieve performance scores of 0.128, 0.615, 0.513, 0.705, 0.346, 0.295, 0.564, and 0.769 in BTCV, and 0.238, 0.664, 0.83, 0.679, and 0.757 in AMOS 2022, respectively. The experimental results demonstrate that Swin unetr exhibits the highest usability and comprehensiveness, while U-Net displays the lowest score. However, the ranking of the remaining models varies with different datasets, with little score repetition. Additionally, when changing the threshold for evaluation, the change of the final score is smoother than that of the Usable Region Estimate, which indicates that the dependence of the algorithm on sample size is reduced. Meanwhile, MIMO evaluates multiple correctness indicators, confidence estimates, and multi-organ using a unified metric, providing comprehensive performance evaluation insights not fully described by overall correctness, confidence calibration errors, and Usable Region Estimate. Finally, it presents a concise result suitable for doctors to compare models and engage in clinical deployment.

\begin{figure}[htbp]
  \centering
  \subfigure{\includegraphics[width=0.45\textwidth]{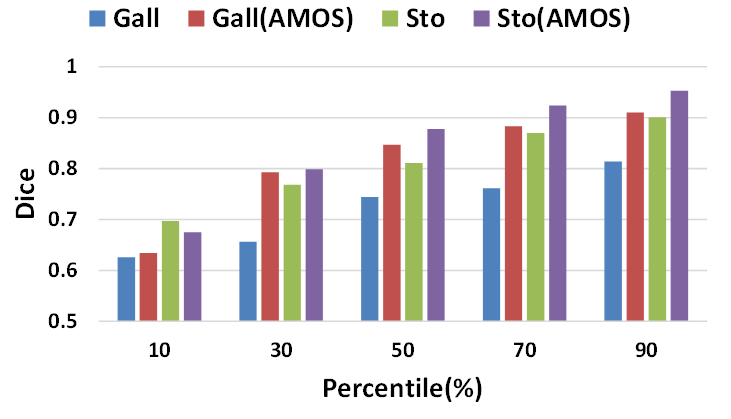}}
  \subfigure{\includegraphics[width=0.45\textwidth]{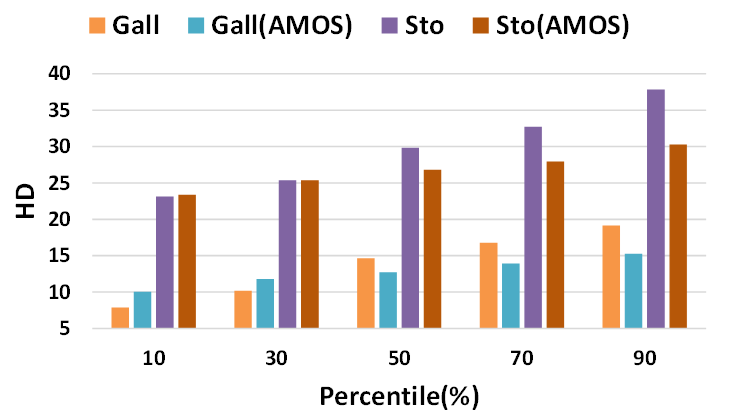}}
  \caption{Take the gallbladder and stomach as examples, the threshold of correctness scores under different percentiles. \textbf{Left}:The variation of the Dice coefficient. \textbf{Right}:The variation of Hausdorff distance. Note: Gall and Sto represent gallbladder and stomach, respectively}
  \label{Fig.5}
\end{figure}

\begin{figure}[htbp]
  \centering
  \subfigure{\includegraphics[width=0.8\textwidth]{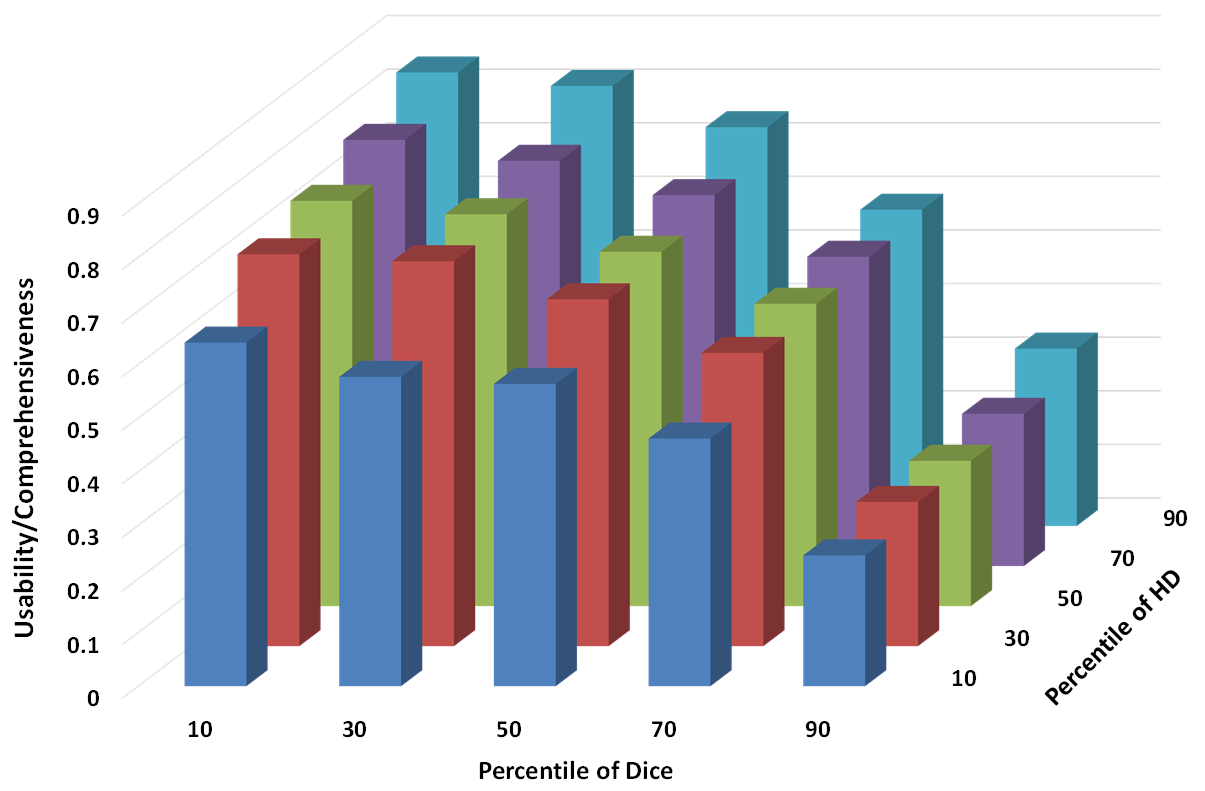}}
  \caption{The distribution of usability and comprehensiveness of Attention u-net in different percentiles in BTCV dataset.}
  \label{Fig.6}
\end{figure}

\noindent \textbf{Others.} The generation of thresholds is a crucial step in evaluating the performance of models. As depicted in Fig.\ref{Fig.5}, the difference between the maximum and minimum thresholds of Dice coefficient and Hausdorff distance can reach 0.278 and 14.71 respectively. The selection of different percentiles yields different results, with lower percentile values leading to lower thresholds, and vice versa. Consequently, the requirements for the Dice coefficient and Hausdorff distance indices vary accordingly. The usability and comprehensiveness score is impacted by changes in these requirements, as depicted in Fig.\ref{Fig.6}, with a decrease in the Dice coefficient requirement leading to an increase in the final score, and an increase in the Hausdorff distance requirement leading to a decrease in the score. In addition, different emphases on different indicators will lead to an inconsistent ranking of models. During the experiment, we observed that Transunet outperforms Swin unetr when the Dice coefficient requirement is low, but its usability drops significantly when the requirement is increased. Thus, a model that can maintain a stable and high usability and comprehensiveness score under different percentiles is deemed more favorable.

\begin{table}[H]
\renewcommand\arraystretch{1.5}
\centering % Label your table accordingly
\caption{\label{tab2}Testing the robustness of MIMO: the frequency (standard deviation) of violating the original ranking results of the model in the new sample (the lower the better)}
\begin{tabular}{ccccc}
\toprule[0.35mm]
\textbf{Models} & \textbf{BTCV(\%)} & \textbf{AMOS 2022(\%)} \\
\midrule[0.15mm]
U-Net & 12.85($\pm$3.99) & 2.8($\pm$1.67)\\
V-Net & 18.2($\pm$2.61) & 10.45($\pm$2.31) \\
nnU-Net & 17.3($\pm$3.57) & 9.85($\pm$2.72) \\
Attention u-net & 19.7($\pm$5.03) & 8.7($\pm$2.75) \\
Transunet & 17.85($\pm$2.81) & 11.2($\pm$3.17) \\
Swin-unet & 13.75($\pm$2.92) & 8.05($\pm$2.5) \\
Unetr & 18.6($\pm$3.62) & 9.2($\pm$2.01) \\
Swin unetr & 17.55($\pm$4.88) & 8.55($\pm$3.05) \\
\bottomrule[0.35mm]
\end{tabular}
\end{table}

\begin{table}[H]
\renewcommand\arraystretch{1.5}
\centering % Label your table accordingly
\caption{\label{tab3}Bootstrapping is not used in MIMO and repeat the experiment: the frequency (standard deviation) of violating the original ranking results of the model}
\begin{tabular}{ccccc}
\toprule[0.35mm]
\textbf{Models} & \textbf{BTCV(\%)} & \textbf{AMOS 2022(\%)} \\
\midrule[0.15mm]
U-Net & 32.75($\pm$3.85) & 5($\pm$1.52)\\
V-Net & 50.05($\pm$3.03) & 46.85($\pm$5.91)\\
nnU-Net & 44.6($\pm$5.38) & 41.75($\pm$6.45) \\
Attention u-net & 42.1($\pm$5.45) & 38.85($\pm$6.37) \\
Transunet & 24.4($\pm$3.66) & 47($\pm$5.69) \\
Swin-unet & 33.55($\pm$4.78) & 32.35($\pm$6.94) \\
Unetr & 32.15($\pm$3.45) & 45($\pm$5.3) \\
Swin unetr & 29.05($\pm$4.2) & 48.2($\pm$4.97) \\
\bottomrule[0.35mm]
\end{tabular}
\end{table}

\subsection{Robustness Experiment}

 In this study, we evaluate the robustness of our proposed method on unseen samples. Our approach involves randomly splitting the original test sets into two halves. The first half is utilized to calculate the threshold for subsequent screening of qualified or unqualified organ data, while the other half is used to test and calculate the final usability and comprehensiveness score of each model. The experiment is repeated 100 times, and the frequency of inconsistent model ranking with the original test results is reported as an indicator of error frequency and robustness of the proposed method. We repeat the entire procedure 20 times to report the mean and standard deviation of the error frequency. In addition, we remove the Bootstrapping section from MIMO and repeat the experiment. Table\ref{tab2} indicates that our proposed method provides a reliable estimate of model availability and comprehensive performance. Table\ref{tab3} shows the frequency of violateing the original ranking results increases more than that in Table\ref{tab2}, and it further confirms the necessity of using Bootstrapping to calculate confidence intervals within MIMO.

\subsection{Discussion}
 In the preceding experiments, our proposed method demonstrates low error frequency, highlights its utility in multi-indicator and multi-organ segmentation evaluation. However, it is worth noting that all data have labels in our experiments. It is a direction to be explored how to evaluate the unsupervised medical image segmentation models. Additionally, how to effectively involve doctors and experts in the process of model evaluation is also a critical problem to be solved.

\section{Conclusion}
  Starting from the application scenario of multi-indicator and multi-organ medical image segmentation, this paper introduces a method suitable for evaluating multiple models, dubbed MIMO. Through the way of autonomously generating thresholds within the model, this method sorts and filters the multi-organ sample data according to the correlation between the predictive correctness index and the confidence estimation value. Finally, the segmentation results of each organ above or below the threshold are distinguished in a set manner, and the usability and comprehensiveness score of the model is visualized.

  Beyond The Cranial Vault (BTCV) and Multi-Modality Abdominal Multi-Organ Segmentation Challenge (AMOS 2022) were used to test the reproduced models and evaluate them with the proposed method. The experimental results are analyzed from four perspectives: correctness, confidence estimation, Usable Region, and the proposed method. Additionally, robustness experiments were conducted on MIMO, demonstrating a low error frequency. Finally, MIMO provides novel insights and a simple and intuitive measurement standard for  evaluating multi-index and multi-organ medical image segmentation.

\newpage

%\bibliographystyle{johd}
%\bibliography{reference}
\begin{refcontext}[sorting = none]
\printbibliography
\end{refcontext}

\end{document}